\documentclass[aps,twocolumn]{revtex4}

\usepackage{graphicx}
\usepackage{color}
\usepackage{subfig}

\begin{document}

\title{A detection pipeline for galactic binaries in LISA data}
\author{\surname {Tyson} B. Littenberg}\affiliation{Maryland Center for Fundamental Physics, Department of Physics, University of Maryland, College Park, MD  20742}\affiliation{Gravitational Astrophysics Laboratory, NASA Goddard Spaceflight Center, 8800 Greenbelt Rd., Greenbelt, MD  20771}
\date{\today}

\begin{abstract}
The Galaxy is suspected to contain hundreds of millions of binary white dwarf systems, a large fraction of which will have sufficiently small orbital period to emit gravitational radiation in band for space-based gravitational wave detectors such as the Laser Interferometer Space Antenna (LISA).  LISA's main science goal is the detection of cosmological events (supermassive black hole mergers, etc.) however the gravitational signal from the galaxy will be the dominant contribution to the data -- including instrumental noise -- over approximately two decades in frequency.  The catalogue of detectable binary systems will serve as an unparalleled means of studying the Galaxy.  Furthermore, to maximize the scientific return from the mission, the data must be ``cleansed'' of the galactic foreground.  We will present an algorithm that can accurately resolve and subtract $\gtrsim 10000$ of these sources from simulated data supplied by the Mock LISA Data Challenge Task Force.  Using the time evolution of the 
gravitational wave frequency, we will reconstruct the position of the recovered binaries and show how LISA will sample the 
entire compact binary population in the Galaxy. 
\end{abstract}

\maketitle

\section{Introduction}
\label{intro}
With the direct detection of gravitational waves (GWs) poised to be made this decade, the long-awaited dawn of gravitational wave astronomy is near.
%Just as traditional photon astronomy has profited from accessing the entire electromagnetic spectrum, 
%so to would gravitational wave (GW) astronomy benefit from detectors covering many decades in frequency.  
Similar to traditional photon astronomy, different bands of the GW spectrum offer unique channels of information about the Universe.
%With existing GW observatories, 
High-frequency gravitational waves are the target of ground based interferometers, and the LIGO/Virgo collaboration~\cite{LSC,Virgo} have developed astonishingly sensitive observatories which are poised to bring GW astronomy out of its infancy.

While small wavelength gravitational waves are, observationally, the most readily accessible, the richest signal space in the GW universe is at frequencies far too low for Earth-based instruments.  A space-borne interferometer is the only foreseeable way of reaping the bounty of information transmitted at longer wavelengths, and allowing gravitational wave astronomy to reach its full potential.

Unique to the mHz frequency range is the existence of known sources, comprised of close binary star systems in the Galaxy, mostly the AM CVn-type binary stars~\cite{Stroeer:2006rx}.  These individual objects, discovered electromagnetically, are the near-by representatives of a much larger population of low frequency gravitational wave sources on our cosmic doorstep.  Population synthesis models for the galaxy predict some 60 \emph{million} binary star systems emitting gravitational radiation at frequencies between 0.1 and 10 mHz~\cite{Hils:1990vc, Nelemans:2001hp, Nelemans:2003ha}.  Because gravitational waves interact very weakly with intervening matter, a staggeringly large number of these objects, distributed throughout the entire Galaxy, are potentially observable.

Maximizing what can be gleaned from mHz GW data will require solving unique analysis problems in gravitational wave astronomy.  In particular, the challenges posed by these galactic binaries are the sheer number of sources, and the degree with which they overlap one another in signal space, smearing together to form a confused blend of gravitational wave power which, depending on details of the detector, can be orders of magnitude larger than the instrumental noise floor.  The total number of binaries which are individually resolvable out of this population is unknown and poses a very large-dimension model selection problem.  The harm in over-fitting the data, (i.e, tolerating large false alarm probabilities) or under-fitting the data (accepting large false dismissal rates) not only affects the science that can be done with the catalogue of resolved signals, but can also impact the data analysis efforts for more distant sources of gravitational radiation which share the same bandwidth.  
%This includes the mergers of supermassive black-holes and the capture of compact objects by supermassive black holes.

To prepare for this challenge, we have set out to build a ``detection pipeline'' which can automatically solve every facet of the galactic binary detection problem.  To wit, we need to locate candidate sources in the data, select for the most parsimonious number, accurately estimate the physical parameters that describe the system, and cleanly regress the sources from the data.  While the analysis software we have built is flexible with regards to the details of the instrument, we use the Laser Interferometer Space Antenna (LISA), a joint NASA/ESA mission concept~\cite{LISA}.  This choice was driven by our intention 
of participating in round four of the Mock LISA Data Challenges (MLDC)~\cite{MLDC}.  MLDC datasets are released in pairs:  one coming with the list of signal parameters within the data, and one without (henceforth, the ``training data'' and ``blind data'', respectively).  For this study, we will focus on analyzing the training data in which all other types of sources have been removed, leaving behind only the galactic binaries and the instrument noise.  The capabilities of the algorithm on this reduced dataset will serve as a realistic demonstration of what we could achieve on the blind data, as the only cosmological sources which can impact the number of resolvable binaries are the brightest of the binary black hole mergers and cosmic strings, neither of which pose a serious challenge to existing search algorithms.

The challenge presented by the galactic foreground has been addressed with iteratively more sophistication in past  MLDCs~\cite{Babak:2008sn}.  Challenge 2 was the first to simulate a complete galaxy and in response to this, Crowder and Cornish developed the {\tt BAM} algorithm~\cite{Crowder:2006eu} which, to date, has been the most successful demonstration of galactic binary data analysis.  
The {\tt BAM} codes themselves, which have been lost to the sands of time, did not model the evolution of a binary's orbital period (nor did Challenge 2).  Other existing algorithms which have worn their teeth on the MLDC data sets include an F-statistic maximization scheme using a Nelder-Mead simplex algorithm~\cite{Blaut:2009si}, a hierarchical cleaning algorithm (also employing the F-statistic)~\cite{Whelan:2009jr} and an MCMC search algorithm featuring a Markovian delayed rejection proposal distribution~\cite{Trias:2009km}.  Beyond MLDC entries there have been numerous proof of principal studies including (but not limited to), Refs.~\cite{Umstatter:2005jd, Littenberg:2009bm, Stroeer:2009hj}.

Our goal for this paper is to build a new analysis pipeline to process the signal from the population of galactic binaries.   We are, in effect, attempting to extend the {\tt BAM} algorithm by including frequency evolution in our model of the waveforms, and incorporating technical advancements made in GW data analysis since the second round of the MLDC.  

The paper is organized as follows:
In \S\ref{analysis} we outline the basics of the data analysis methodology used in this work.
In \S\ref{waveforms} we describe our model for the data, both for the GW signals as well as the instrument noise.
The search algorithm is spelled out in \S\ref{algorithm}, and results from the MLDC challenge 4 training data are discussed in \S\ref{catalogue}.
We close in \S\ref{discussion} with discussion of future work which will utilize the bright source catalogue as a novel tool for Galactic astronomy. 

\section{Data analysis basics}
\label{analysis}
There are two desired products of the data analysis procedure.  For each model of the data $(\mathcal{M})$ under consideration, we need to find the ``best fit'' model parameters and have some sense of how well these parameters are constrained, as well as determine which model is most strongly supported by the data.

The mathematical foundation on which the data analysis theory is built is Bayes' theorem which, when cast as a tool for solving inference problems, takes the form
\begin{equation}\label{bayes}
p(\vec{\theta}|s,\mathcal{M}) = \frac{p(s|\vec{\theta},\mathcal{M})p(\vec{\theta}|\mathcal{M})}{p(s|\mathcal{M})}.
\end{equation}
The left hand side of equation~\ref{bayes} is the posterior distribution function (or ``posterior'' for short) for parameters $\vec{\theta}$ given the data $s$, from which parameter estimation conclusions can be drawn within model $\mathcal{M}$ (see, for instance, ~\cite{Gregory} for a thorough introduction).  The numerator of the right hand side contains the product of the likelihood $p(s|\vec{\theta},\mathcal{M})$ -- our ``goodness of fit'' measure -- and the prior, encoding our knowledge before the new data were collected.  The denominator is the evidence, or marginalized likelihood for $\mathcal{M}$.  Comparisons of $p(s|\mathcal{M})$ between different models reveal which is most strongly supported by the data and, assuming uninformative priors on the models themselves, should be taken as the preferred representation.

In the Bayesian framework, one needs only to define the likelihood and prior distributions, and the rest of the analysis is reduced to an oft time-consuming calculation.  We prefer the Markov Chain Monte Carlo (MCMC) family of algorithms to perform the calculations, although Nested Sampling, and its offspring {\tt MultiNest}, have been used to similar affect (Refs.~\cite{Veitch:2008ur,Feroz:2009de}).  There is no shortage of data analysis literature describing the concept of MCMCs.  However, for the sake of introducing vocabulary and notation, we briefly sketch the algorithm.  

The first ``link" in the Markov chain is some random position in parameter space $\vec{\theta}_x$ for which we have evaluated the likelihood $p(s|\vec{\theta}_x,\mathcal{M})$ and prior probability $p(\vec{\theta}_x|\mathcal{M})$ .  From there, a new trial position, $\vec{\theta}_y$, is drawn from a proposal distribution $q(\vec{\theta}_y|\vec{\theta}_x)$.  The likelihood and prior probability are evaluated for $\vec{\theta}_y$, and it is adopted as the next sample in the chain with probability ${\alpha}_{\vec{\theta}_x\leftrightarrow \vec{\theta}_y} =\text{min}[1,H_{\vec{\theta}_x\leftrightarrow \vec{\theta}_y} ]$ where $H_{\vec{\theta}_x\leftrightarrow \vec{\theta}_y} $ is the Hastings ratio
\begin{equation}\label{Hastings}
H_{\vec{\theta}_x \rightarrow \vec{\theta}_y}=\frac{ p(s|\vec{\theta}_y,\mathcal{M}) p(\vec{\theta}_y|\mathcal{M}) q(\vec{\theta}_x|\vec{\theta}_y) }{ p(s|\vec{\theta}_x,\mathcal{M}) p(\vec{\theta}_x|\mathcal{M}) q(\vec{\theta}_y|\vec{\theta}_x) }.
\end{equation}
If $\vec{\theta}_y$ is rejected, the chain remains at $\vec{\theta}_x$ and a new trial position is considered.  This process of stochastically stepping through parameter space repeats until some convergence criteria are satisfied.  Equation~\ref{Hastings} is derived from the detailed balance condition, which is satisfied if the probability of being at state $\vec{\theta}_x$ and transitioning to state $\vec{\theta}_y$ is the same as being at $\vec{\theta}_y$ and moving to $\vec{\theta}_x$.  By fulfilling this condition when adopting new solutions in the chain, the number of iterations spent in a particular region of parameter space, normalized by the total number of steps in the chain, yields the probability that the model parameters have values within that region.

The choice of $q(\vec{\theta}_y|\vec{\theta}_x)$, by construction, does not alter the recovered posterior distribution function. The proposal distribution does, however, dramatically affect the acceptance rate of trial locations in parameter space and, therefore, the number of iterations required to satisfactorily sample the posterior.  Considering Eq.~\cite{Hastings}, and ignoring the proposal distribution ratio, the algorithm will accept moves to positions in parameter space which have higher posterior weight like any good ``hill-climbing'' search algorithm.  The novelty of the MCMC is its willingness to adopt a worse fit solution, mitigating the potential of getting trapped by local features of the posterior.  Despite this built in attribute of the algorithm, extremely multimodal distributions can be a detriment to the efficiency of the sampler.  The posteriors for typical gravitational wave sources have been found to exhibit substantial sub-dominant modes which can be too difficult for a straight-forward MCMC to overcome in a reasonable amount of time.  To help alleviate the challenge posed by these distributions, we include parallel tempering~\cite{Swendsen:1986} -- a set of chains, running simultaneously, each at a higher ``temperature'' -- as is becoming standard in GW applications of MCMCs, e.g.~\cite{vanderSluys:2008qx,Littenberg:2010gf,Cornish:2011ys}.  

For the model selection facet of the problem, we use a trans-dimensional (or ``reverse jump'') Markov Chain Monte Carlo (RJMCMC)~\cite{Green:1995,Green:2003} as the tool for sampling the target posterior distribution function in model space.  The RJMCMC stochastically moves between models while satisfying detailed balance, so the number of iterations the chain spends in a particular model is proportional to the marginalized likelihood for that model.  This class of MCMC algorithm has previously been used to study the galactic binary detection problem on toy problems~\cite{Umstatter:2005su} and smaller data sets~\cite{Littenberg:2009bm, Stroeer:2009hj}, but has yet been turned loose on data containing a complete simulation of the GW signals from the entire Galaxy. 

The MCMC algorithm provides the machinery to employ Bayes' theorem to our data analysis problem.  Properly interpreting the results from the MCMC requires detailed understanding of our model for the data, particularly the definition of the likelihood function.  In the following section we will explicitly spell 
out our construction for the data, including the waveform and noise models, the likelihood function, and the priors used in the analysis.

\section{The data model}
\label{waveforms}
We model the data $s$ as having two contributions:
\begin{equation}
\tilde{s}_{\kappa} = \tilde{n}_{\kappa}(\vec{\eta}_{\kappa}) + \sum_i^M \tilde{h}^i_{\kappa}(\vec{\lambda}^i).
\end{equation}
The instrument noise, $n$, is assumed to be stationary and gaussian with colored spectral density parameterized by $\vec{\eta}_{\kappa}$.  The gravitational wave component of the data is the superposition of $M$ gravitational wave templates in, each parameterized by $\vec{\lambda}$ which will be described in detail later.  The subscript $\kappa$ denotes the different interferometer channels synthesized from the LISA phase-meter data.  We use the usual noise orthogonal AET channels~\cite{Tinto:2002de}.  Because we expect all of the galactic binaries to be well below the LISA transfer frequency $f_* = c / 2 \pi L$ we can neglect the T channel which is, in effect, GW-free for $f<f_*$.

\subsubsection{The waveform model}
Galactic binaries in the LISA band are expected to exhibit relatively little frequency evolution during the lifetime of the mission.  Thus, the phase of the GWs emitted from the binary can be safely approximated as $\Phi(t) = \varphi_0 + 2\pi f_0 t  + \pi \dot{f}_0 t^2 + ...$ where higher order derivatives of $f$ can be neglected for binaries below $\sim9$ mHz during a five-year-long mission ~\cite{Stroeer:2005cv}.  For this work, we set $\ddot{f}_0$ and all higher derivatives of $f$ to $0$, as is the case in the MLDC data simulations.

Given these assumptions about the phase evolution of the binary, as well as restricting the templates to circular orbits (perhaps a dubious constraint, see~Ref. \cite{Willems:2007nq}), We can fully describe a GB waveform with eight parameters:
\begin{equation}
\vec{\lambda}\rightarrow\left(f_0,\dot{f}_0,\theta,\phi,\mathcal{A}_0,\iota,\psi,\varphi_0\right)
\end{equation}
where the subscript $0$ indicates the value taken at the first time-sample in the data.  Parameters $\left\{\theta,\phi\right\}$ describe the sky-location of the binary in ecliptic coordinates, and $\left\{\iota,\psi,\varphi_0\right\}$ are angles that fix the orientation of the binary.  The amplitude
\begin{equation}\label{Amp}
A_0 = \frac{2\mathcal{M}^{5/3}\left(\pi f_0\right)^{2/3}}{D_L}
\end{equation}
couples the chirp mass $\mathcal{M}$ and luminosity distance $D_L$ preventing the independent measurement of either quantity.  If the binary orbital evolution is driven purely by the emission of gravitational waves (as opposed to, for instance, mass transfer between the individual stars in the binary), then the linear term in the frequency evolution depends only on the frequency and chirp mass via:
\begin{equation}\label{fdot}
\dot{f} = \frac{96}{5}\pi^{8/3}\mathcal{M}^{5/3}f^{11/3}.
\end{equation}
Sources which satisfy this condition will henceforth be referred to as \emph{detached} binaries.  For this data set, we have the advantage of knowing that any binaries with $\dot{f}_0 > 0$ are detached, allowing us to make a determination of $D_L$.  When analyzing data from the Galaxy itself, we will not have this foresight.  For the rare cases where $\ddot{f}_0$ can also be measured, systems being driven only by the emission of gravitational radiation must satisfy the braking index condition:
\begin{equation}
n\equiv\frac{f \ddot{f}}{\dot{f}^2}.
\end{equation}
LISA's ability to measure $\ddot{f}_0$, and how ignoring this parameter impacts the data analysis, has been preliminarily explored in \cite{Stroeer:2005cv} and~\cite{Stroeer:2009uy}.  We will address this important detail in a follow-on study~\cite{Larson}.

To compute the instrument response to a particular galactic binary signal we use the fast-slow decomposition as detailed in the appendix of~\cite{Cornish:2007if}.

We utilize a prior on the location of any given binary constructed from the number density of stars in the galaxy.  Our model for the galactic distribution is similar to that from which the MLDC datasets were drawn~\cite{Nelemans:2001hp,Nelemans:2003ha}.  The density profile has two components, one from the disk and one from the bulge:
\begin{eqnarray} \label{galaxy_density}
\rho_{\rm bulge}&=&  \frac{1}{ \left( \sqrt{\pi} R_b\right)^3 } e^{ \frac{-r_{\rm gc}^2}{R_b^2} }  \nonumber \\
\rho_{\rm disk}&=& \frac{1}{ 4\pi R_d^2 Z_d }  {\rm sech}^2\left(\frac{z_{\rm gc}}{Z_d}\right)e^{\frac{-u_{\rm gc}}{R_d}}\nonumber \\
\rho_{\rm galaxy} &=& A\rho_{\rm bulge} + (1-A)\rho_{\rm disk}
\end{eqnarray}
where $r_{\rm gc}^2 = x_{\rm gc}^2 + y_{\rm gc}^2 + z_{\rm gc}^2$, $u_{\rm gc}^2~=~x_{\rm gc}^2 + y_{\rm gc}^2$, and $\{x_{\rm gc},\ y_{\rm gc},\ z_{\rm gc}\}$ are the cartesian galactic coordinates of the source.  For our purposes, the parameters of the galaxy distribution used are~\cite{Adams_private}
\begin{equation}
\left\{R_b,R_d,Z_d,A\right\} = \left\{690\ {\rm pc}, 2520\ {\rm pc},302\ {\rm pc},0.24\right\}\nonumber
\end{equation}
Using this distribution we build a joint prior
\begin{equation}\label{joint_prior}
p(f_0,\dot{f}_0,\theta,\phi,\mathcal{A}_0) = C\rho_{\rm galaxy}
\end{equation}
with a complicated normalization constant $C$ that we approximate by Monte Carlo integration over the prior volume.  The quantities not constrained by this prior are the orientation parameters, which we take as having uniform \emph{a priori} distributions over $[0,2\pi]$ for $\psi$ and $\varphi_0$, and $[-1,1]$ for $\cos{\iota}$.

To utilize the prior in Eq.~\ref{joint_prior} for templates of detached systems we determine the distance to the binary using Eqs.~\ref{Amp} and~\ref{fdot}.  For mass-transferring binaries we have no way of making such a determination without better understanding of the orbital dynamics.  We construct a separate prior for such systems, marginalizing Eq.~\ref{joint_prior} over $f_0$, $\dot{f}_0$, and $\mathcal{A}_0$.  We are left with a prior on $\{\theta,\phi\}$ only, and adopt uniform distributions on the marginalized parameters.
%Equation~\ref{galaxy_density} is, as priors go, a particularly strong constraint.  However, the distribution in Eq.\ref{galaxy_density} is motivated by observations and is thus a reasonable snap-shot of our prior state of knowledge.  There are, of course, competing descriptions of the macroscopic properties for our galaxy, and these differences may have a significant impact on the performance of our algorithm.  We spot-checked the search in different frequency ranges using uniform priors on all parameters and saw measurable, but not catastrophic, differences in the results. {\color{red} Can't get away with this w/out some examples}.  
\subsubsection{The noise model}

Given nominal levels for the shot- and acceleration-noise ($S_s$ and $S_a$), the baseline noise power spectral density for the LISA $A$ and $E$ channels is
\begin{eqnarray}
S_n(f) &=& \frac{4}{3}\sin^2\frac{f}{f_*}\left[  \left(2+\cos \frac{f}{f_*}\right)S_{s} \right.\nonumber \\
&+& \left. 2\left(3+2\cos \frac{f}{f_*}+\cos \frac{2f}{f_*}\right)\frac{S_{a}}{(2\pi f)^4} \right].
\end{eqnarray}
To this we must add an estimate of the confusion noise $S_c$ which is derived from data simulations
\begin{equation}
S_{c}(f) = \left\{ \begin{array}{ll}
         10^{-44.8}f^{-2.4}  & 10^{-4} < f < 4.5\times10^{-4}\\
         10^{-47.15}f^{-3.1} & 4.5\times10^{-4} < f <1.1\times10^{-3}\\
         10^{-51}f^{-4.4} & 1.1\times10^{-3} < f <1.7\times10^{-3}\\
         10^{-74.7}f^{-13} & 1.7\times10^{-3} < f <2.5\times10^{-3}\\
         10^{-59.15}f^{-7} & 2.5\times10^{-3} < f <4\times10^{-3}\\
         \end{array} \right.
\end{equation}

To allow for modeling error in the noise levels, and the vagaries of the particular noise realization in the data, we include parameters which characterize departures from this theoretical noise power spectral density as described in~\cite{Littenberg:2009bm}.  
%In this prescription, the noise level over the $i^{\rm th}$ sub-segment of the data comprised of $N_{NB}$ bins is rescaled 
%as $S_n(f) \rightarrow \eta_i S_n(f)$.  
A separate noise level is defined for each of several narrow bandwidth segments of data, each of length $N_{\rm NB}$ frequency bins.  The i$^{\rm th}$ 
segment is rescaled as $S_n(f) \rightarrow \eta_i S_n(f)$.
For Gaussian noise, the expectation value for $\eta$ when measured over $N_{\rm NB}$ bins is $\sigma^2_{\eta} = 1/\sqrt{N_{\rm NB}}$.  We accommodate for additional ignorance with respect to the noise level by using a normal distribution $N[1,4\sigma^2_{\eta}]$ as the prior on each $\eta_i$.

\subsubsection{The likelihood function}
With the noise and signal models now declared, the likelihood $p(s|\vec{\theta})$ is computed over $N$ Fourier bins, built from the assumption of colored Gaussian noise where the noise power spectral density is being fit over $N_{\rm seg} = N/N_{\rm NB}$ narrowband segments of data, via:
\begin{equation}\label{likelihood}
\ln p(s|\vec{\theta}) = -\frac{1}{2}\sum_{\kappa}^{\rm{A,E}}\left[ \left( r_{\kappa}|r_{\kappa}\right) + N_{\rm NB}\sum_n^{N_{\rm seg}} \ln\eta_{\kappa}^n \right].
\end{equation}
The residual 
\begin{equation}
r_{\kappa} = s_{\kappa}-\sum_i^M h^i_{\kappa}(\vec{\lambda}^i)
\end{equation}
appears in equation~\ref{likelihood} inside of the noise-weighted inner product defined as
\begin{equation}
(a|b) = \frac{2}{T_{\rm obs}}\sum_f\frac{\tilde{a}^*(f) \tilde{b}(f) + \tilde{a}(f)\tilde{b}^*(f)}{\eta(f) S_n(f)}
\end{equation}
where, as described above, $\eta$ takes the same value over $N_{\rm NB}$ Fourier bins.

\section{The %{\tt BAM+} 
algorithm} 
\label{algorithm}
\subsection{Overview}
In this section we will describe each phase in the pipeline.  Following a coarse overview of the entire procedure, a detailed, step-by-step description of the algorithm can be found in the subsections.  

The data are first divided into small bandwidth subsets, each of which is independently analyzed.  
In each window, the analysis goes through roughly three phases:
\begin{itemize}
\item \emph{Search}:  To locate the regions of high probability in parameter space.
\item \emph{Characterization}:  To globally sample the posterior distribution for the model parameters (parameter estimation). 
\item \emph{Evaluation}:  To calculate model evidence and determine which of the models are favored (model selection).
\end{itemize}
In this paper, the Search phase is subdivided into a ``burn-in'' phase where we use a reduced parameter space to look for modes in the distribution, 
and a full parameter search that is tuned to move between the modes in order to identify the region in parameter space encompassing the global maximum 
likelihood.

All of our Markov chain runs use a variety of proposal distributions, from uniform proposals over the full prior volume, to small jumps along the eigenvectors 
of the covariance matrix estimated by the Fisher information matrix~\cite{Littenberg:2009bm}.

Each data window is studied iteratively.  At each iteration, we include an additional template in the model until we reach the maximum evidence.  
We apply a variety of tricks to increase the efficiency during the Search phase, many of which spoil the statistical properties of the chain.  The ``illegal'' search chains are used to produce proposal distributions for the subsequent Characterization/Evaluation phases, where we take care to satisfy detailed balance.
% and sample the correct distribution.  
Parameter estimation and model selection are performed simultaneously, as we include the number of templates in the model as a parameter.  For example, during iteration $I$ the RJMCMC is allowed to move between models containing $0\leq i \leq I$ templates.  The model in which the RJMCMC spends the most iterations, $\mathcal{M}^i_{\rm max}$, is the one with the highest evidence.  If $i_{\rm max}<I$ then that window is finished being analyzed, and the MAP parameters from model $\mathcal{M}^i_{\rm max}$ are stored in the master list for the full dataset.

\subsection{Preparing the Data}
The bandwidth of a typical galactic binary waveform is sufficiently small that the data can be divided into small subsets, or windows, each of which spans a relatively small range in frequency, and each window can be analyzed independently.  
For practical purposes, this makes it simple to scale the analysis code from a testing platform to running on the full dataset.  While many CPUs are required to process all of the data, they do not need to talk to one another while doing so.

The galactic binary search will be hindered at low frequency by power from high SNR black hole binaries in the data, as well as bursts from cosmic strings.  Both the black hole binaries and the cosmic strings have very unique time-frequency characteristics.  This means the brightest sources can be cleanly removed from the data without marring the signal from the galaxy.  The accurate detection of black holes has been a main theme of LISA science studies and is definitely a manageable task, while the high-accuracy detection of the cosmic strings has been successfully demonstrated in the previous round of the MLDC~\cite{Babak:2009cj}.  We do not see the removal of black hole or cosmic string sources as a substantial technical challenge, and so have focused the efforts in this paper on training data with only instrument noise and the signal from the galaxy.  For the blind data analysis to follow this work, a collaborative effort will be needed to perform this first cleaning step.

\begin{figure}[htbp]
   \includegraphics[width=0.5\textwidth]{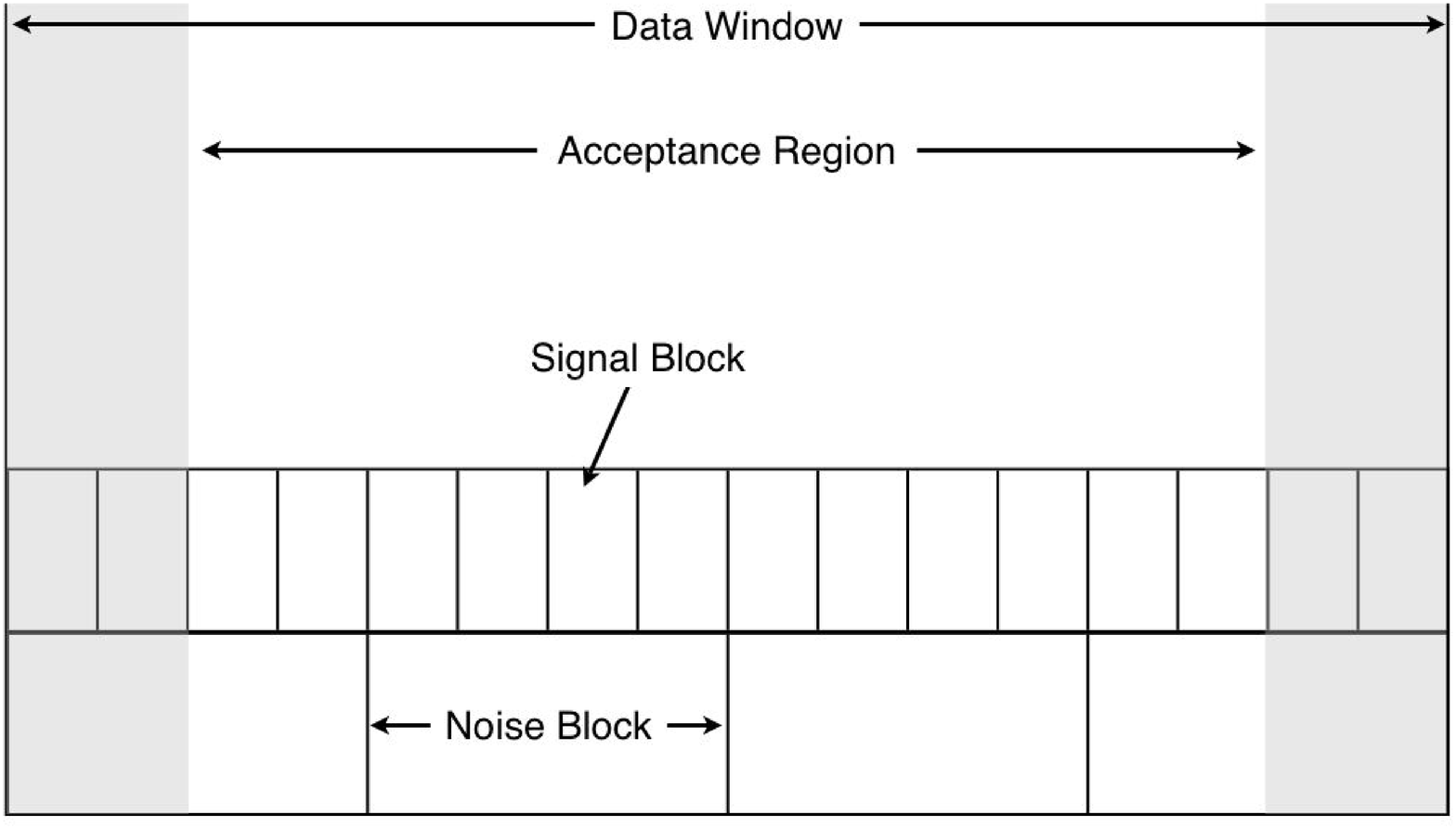} % requires the graphicx package
   \caption{A cartoon depiction of a data window to be independently searched.  The shaded portions are where detected binaries will be excluded from the master list.  All templates within the same Signal Block are updated simultaneously.  One parameter $\eta$ is used to measure the noise PSD over each Noise Block.  This concept was derived for the {\tt BAM} algorithm and adapted from Ref.~\cite{Crowder:2006eu}}.
   \label{algorithm:window}
\end{figure}

In each window, care needs to be taken at the edges as templates will try to fit signals from adjacent data segments which have power ``leaking'' into that which is being analyzed.  Thankfully this was addressed in the {\tt BAM} algorithm by having a smaller acceptance region within each window where the initial frequency must fall if the source is going to be taken as a true detection.  Bordering this acceptance region to the edges of the window are the ``wings'' of the data segment where templates are included in the model but the sources they recover are not stored as detections.  Adjacent windows are tiled so that the end of one acceptance region is the beginning of another.  Thus we have full coverage of the data without any overlap between acceptance regions, and without the risk of double-counting sources that happen to lie at the interface of two data segments.  Figure~\ref{algorithm:window} shows a cartoon depiction of a data window.

The size of the windows is not something that can be fixed for all signals.  The amplitude of a galactic binary waveform scales as $f^{2/3}$, meaning the discrete Fourier transform of a signal typically has significant power across a larger bandwidth as we move to higher frequency data.  Furthermore, for detached binaries (which make up the bulk of the galactic population) $\dot{f}$ scales as $f^{11/3}$ so signals with high initial-frequency typically spread their power over more bins during the course of the observation.  

Because of this frequency-dependent bandwidth, the size of the wings and, for efficiency's sake, the size of the windows, is frequency dependent.  
%Table~\ref{windows} shows the window dimensions as a function of frequency.

\subsection{The Search Phase}
The purpose of the search phase is to rapidly locate the sources in the data (i.e., the modes of the posterior distribution function).  Here we are not concerned with satisfying detailed balance, or producing samples which represent the posterior, but instead are focused on efficiency.  The two most substantial cost saving enhancements come from reducing the dimension of the search by maximizing the likelihood over ``extrinsic parameters'' (orientation and distance) using the F-statistic~\cite{Jaranowski:1998qm, Krolak:2004xp}, and by making the signals a bigger target in the search space through simulated annealing.

Simulated annealing is another common trick performed when using Markov Chain Monte Carlo-like methods to rapidly locate the modes of the distribution.  It works by initially suppressing the influence of the likelihood terms in the Hasting's ratio (Eq~\ref{Hastings}) by ``heating'' the distribution being searched 
\begin{equation}
\frac{p(s|\vec{\theta}_y)}{p(s|\vec{\theta}_x)} \rightarrow \left(\frac{p(s|\vec{\theta}_y)}{p(s|\vec{\theta}_x)}\right)^{\beta}
\end{equation}
with the exponent playing the role of an inverse ``temperature'' $\beta=1/T$ with $0 \leq \beta \leq 1$.
Early iterations of the chain are run at high temperature (small $\beta$) forcing the likelihood ratio term to $\sim1$.  The influence of the likelihood is gradually increased as the search space is gradually ``cooled''  at each iteration $i$ via
\begin{equation}
\beta = \left\{ \begin{array}{ll}
         \left( \frac{1}{T_{\rm max}}\right)T_{\rm max}^{i/\tau}  & 0 \leq i < \tau\\
         1 & i \geq \tau\\
         \end{array} \right.
\end{equation}
until $\beta$ goes to 1 and the chains are sampling the target distribution.

Simulated annealing requires some tuning in order for it to actually improve the search's efficiency.  The adjustable parameters to the annealing scheme, the maximum temperature $T_{\rm max}$ and the cooling time $\tau$, need to be custom suited for each problem.  Conceptually, what we want is for $T_{\rm max}$ to be high enough that chains are able to freely explore the full prior range, while not being so absurdly hot that we spend many iterations with no hope of locking on to any of the signal.  A reasonable rule of thumb is that the effective SNR of the signals as seen by the tempered chain is reduced from the true SNR by a factor of $\sim1/\sqrt{T}$.  To this end, we want $T_{\rm max}$ to be $\sim \rm{SNR}^2$, and can reasonably determine it based on the excess Fourier power in the data:  
\begin{equation}
T_{\rm max} =\left(s\vert s\right) - 4N. 
\end{equation}
Setting the cooling rate $\tau$ is an exercise in trial and error, and depends on the bandwidth of the data window, with higher frequencies warranting longer cooling times.

In addition to simulated annealing, and in all other phases of the analysis, we use the now commonplace method of ``parallel tempering''  
where multiple chains are run simultaneously at different temperatures, with exchanges of parameters between chains subject 
to the detailed balance condition.

To further increase the efficiency of the search, we wish to reduce the volume of the search space.  For this purpose, the F-statistic is a tool which has proven extremely useful in LIGO/Virgo searches~\cite{Cutler:2005hc}, and, in a LISA context,  proof-of-principle data analysis black holes~\cite{Cornish:2006ms}, and galactic binaries~\cite{Krolak:2004xp, Crowder:2006eu, Whelan:2009jr}.  Because of it's frequent use in the GW data analysis literature, we leave the details to the aforementioned references. 

While the speed-up in the search time when using the F-statistic is substantial, once the (approximately) best-fit frequency and sky location for each source in the model have been located, it becomes a liability.  For a model containing $N_S$ sources, a single F-statistic evaluation involves $4N_S$ calls to the waveform generator as well as a $4N_S\times4N_S$ matrix inversion, ultimately costing more than $4N_S$ likelihood evaluations.  Therefore, once the F-statistic has done its job and found the modes, the chains are much more efficient reverting to the Gaussian likelihood as described in \S\ref{analysis}.  The points of the chain from the F-statistic search are discarded as the burn-in samples.

At this point we are interested in producing an ensemble of samples that approximate the target posterior distribution function.  However, we are not yet ready to abandon some of  our cost saving measures in favor of detailed balance.  The posteriors for these signals are multimodal and in a few percent of trial runs, the burn-in phase ends on a secondary maximum of the distribution.  The nature of these near-degeneracies was explored with detail in ~\cite{Crowder:2006eu}.  To summarize, the orbital motion of the LISA constellation, as well as the finite number of data points, imparts a harmonic structure to the waveforms on either side of the initial frequency of the binary.  The templates can fix themselves to a sub-dominant harmonic while still achieving overlaps with the injected signals of $\gtrsim70\%$.  

Such features in the posterior expose the weaknesses of an ``out of the box'' MCMC.  While a generic MCMC chain is guaranteed to eventually converge, in most cases we are not willing to wait long enough.  While there are certainly more elegant ways of overcoming these types of challenges, we offer a brute-force approach inspired by delayed rejection~\cite{Tierney:1999, Mira:2001}.  We propose some position in parameter space $\vec{\theta}_y$ and, without asking the Hastings ratio for any input, temporarily adopt this position and evolve the chains from there, searching for a nearby point with higher likelihood.  
After some fixed number of updates the chain arrives at $\vec{\theta}'_y$.  We then look back and calculate the transition probability  $\alpha_{\vec{\theta}'_y\leftrightarrow\vec{\theta}_x}$ using the Hasting's ratio in Equation~\ref{Hastings}.  This transition probability does not satisfy the 
detailed balance condition, and so the samples from the chain will be biased in some way.  For a lot of effort, delayed rejection performs this type of exploration while preserving detailed balance as described in detail within a gravitational wave data analysis context by Trias et al in Ref~\cite{Trias:2009km}.  For our purposes, we are not yet concerned with the statistical properties of our chain and accept the fact that our intermediate distributions will be biased.

Our lazy implementation of delayed rejection is perfectly suited to prevent us from sticking on a sub-dominant harmonic of the waveform.  The secondary modes appear at integer multiples of the LISA modulation frequency $f_m = 1/{\rm year}$.  We propose a shift in frequency by some $nf_m$, where $n$ is a random integer drawn from U[-6,6], adopt that solution, and then allow the chains a few iterations to refine the remaining parameters (in particular, the sky location) before comparing back to the current solution.  This addition to the search procedure dramatically reduces the instances of recovered sources having the central frequency on a harmonic induced by the orbital motion.

\begin{figure}[htbp]
   \includegraphics[angle=270, width=0.5\textwidth]{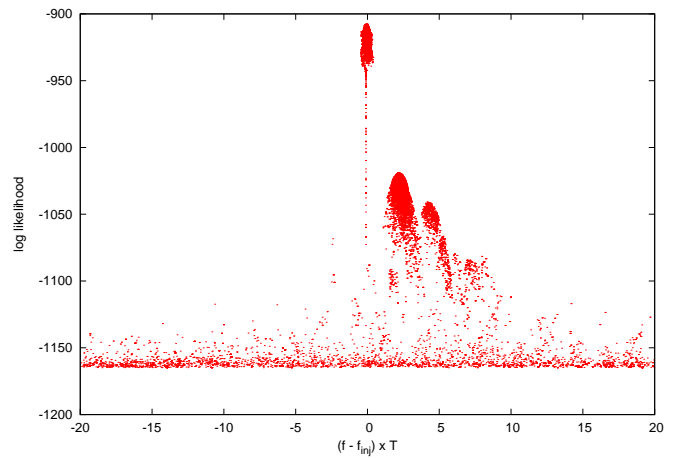} % requires the graphicx package
   \caption{Delayed Rejection at work:  This scatter plot shows samples from a search chain in the $p(s|\vec{\theta}) - f$ plane.  The plot is centered at the true frequency of the injected source.  Delayed rejection enables the chain to efficiently leave a secondary maximum in favor of the true source parameters. }
   \label{Algorithm:DR}
\end{figure}

Figure~\ref{Algorithm:DR} is an instructive depiction of the challenging multimodal structure of the posterior.  We show a scatter plot of chain 
samples in the $p(s|\vec{\theta}) - f$ plane.  The plot is centered at the true frequency of the injected source.  The sub-dominant modes of the distribution are clearly visible as peaks in the likelihood surface occurring at even integer multiples of 
$f_mT = 2$ Fourier bins.  In this example, the search chain approached from higher frequency which is why we only see sub-dominant modes to the 
right of the peak.  In general, these distributions are (roughly) symmetric about the best-fit value of $f_0$.  A single Markov chain without the benefit of delayed rejection or parallel tempering would be severely challenged by these local features of the posterior distribution function.  Missing the global maximum will not only produce poor parameter estimation for that source, but will also leave behind a coherent residual to which subsequent iterations will attempt to fit.  Again, we stress that the (RJ)MCMC algorithm \emph{can} overcome these challenges without any help, but the convergence time is often impractically long.  If chains are not mixing well, the potential for sticking to a local feature of the posterior is increased, and the chances of subsequent iterations rectifying that error during the model selection phase becomes vanishingly small.

\subsection{The Characterization \& Evaluation Phase}
Upon the completion of the search phase, the chains have produced samples of some biased distribution function, the modes of which closely correspond to that of the target posterior.  To accurately refine the characterization of the model and to evaluate wether or not it is the one favored by the data, we run a RJMCMC following all of the rules to ensure the chain 
samples are representative of the target posterior.  RJMCMCs are notoriously difficult to work with in high-dimension problems.  The proposal distributions have to be informative enough that drawing a random point out of (for this example) an 8 dimensional parameter space produces a reasonable fit to the data, but are not so constraining that the improved fit to the data is overly penalized by how strongly you forced the trial solution to that point.  For these techniques to work efficiently, the chains can not tolerate a proposal that is anything but a good approximation to the posterior.

Fortunately for us, we spent the search phase producing such an approximation.  While these biased chains could not be used for model selection or parameter estimation, there are no rules against employing them to build suitable proposal distributions.  This concept was originally suggested by Green~\cite{Green:2003} and we have described in detail the procedure applied to this work in Ref.~\cite{Littenberg:2009bm}.  In short, we bin the chains into an 8D grid using fisher matrix estimates of the parameter variances to set the cell size in the grid.  The number of samples from the chain that land in a given cell is proportional to the probability density in that cell.  The proposal distribution randomly chooses a cell weighted by its probability, and then uniformly draws within the cell to come up with the parameters.  Recently, Farr and Mandel~\cite{Farr:2011sk} introduced an improvement on how to bin the chain samples by using a kD-tree data structure instead of our Fisher-scaled grid.  It is clear that this will further improve the efficiency of these proposals by eliminating the dependence on the grid size, something which we had to carefully tune.  We will transition to the kD-tree decomposition in future work.

As mentioned previously, this analysis is performed iteratively, including an additional signal in the model at each iteration.  During the characterization phase of the procedure on iteration $I$ the RJMCMC is exploring models containing anywhere between 0 and $I$ signals.  The model with the strongest support is the one in which the RJMCMC spends the most iterations.  When two models are similarly supported by the data a more subtle selection has to be made.  Between any two models we can calculate the Bayes factor
\begin{equation}
B_{ij} = \frac{p(s|\mathcal{M}_j)}{p(s|\mathcal{M}_i)} = \frac{{\rm \#\ of\ iterations\ in\ }\mathcal{M}_j}{{\rm \#\ of\ iterations\ in\ }\mathcal{M}_i}
\end{equation}
which is easily interpreted as the odds (ignoring prior preferences for one model over another) that $\mathcal{M}_j$ is preferred over $\mathcal{M}_i$.  Thus, $B_{ij}\sim 1$ means that both models are similarly supported by the data.  While that is a nice interpretation, when assembling a catalogue of detectable galactic binaries and producing a residual fit for subsequent searches, decisions need to be made about which model to pick in these marginally distinguishable cases.  This, sadly, somewhat reduces the Bayes factor to a statistic used for model selection.  Nevertheless, we have to draw a line in the sand, and have chosen for this study a Bayes factor ``threshold'' of 12:1 needed to prefer a higher dimensional model, above which the support for $\mathcal{M}_j$ is canonically considered ``strong''~\cite{Raftery:1996}.  A more satisfying thing to do would be to repeat the analysis for several different $B_{ij}$ cutoffs, producing appendices to the final source catalogue of more speculative detections.

\begin{figure}[htbp]
   \includegraphics[angle=270, width=0.5\textwidth]{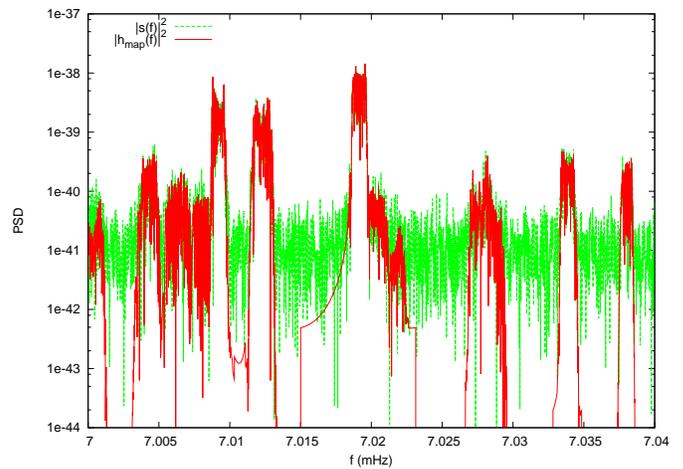} % requires the graphicx package
   \caption{Power spectral density of a small bandwidth segment of the training data.  This figure spans $\sim5$ search 
   windows.  The green [dashed] line is the original data, while the red [solid] line shows the best fit signal model.}
   \label{Algorithm:data}
\end{figure}

If the highest dimension model is supported by the data, we store the map parameters and begin another iteration.  If not, the MAP parameters from the winning model are stored and no more iterations on that window are performed.  Only sources with initial frequency inside the acceptance region are added to the ``master'' list of detections.  Figure~\ref{Algorithm:data} shows $\sim5$ windows' worth of training data, with the best fit waveforms over-plotted.  A relatively high frequency window was chosen for this demonstration so that the different signals in the fit could be distinguished.

\subsection{Source Subtraction}
The biggest cost to performing these MCMC runs is the waveform calculation.  While our waveform model is extremely efficient, there is not much that can be done about the colossal number of templates that need to be computed to resolve 10000+ sources.  To help mitigate this expense, we want to hold fixed the waveform parameters for sources which are not significantly overlapping other sources in the window.  

\begin{figure}[htbp]
   \includegraphics[angle=270, width=0.5\textwidth]{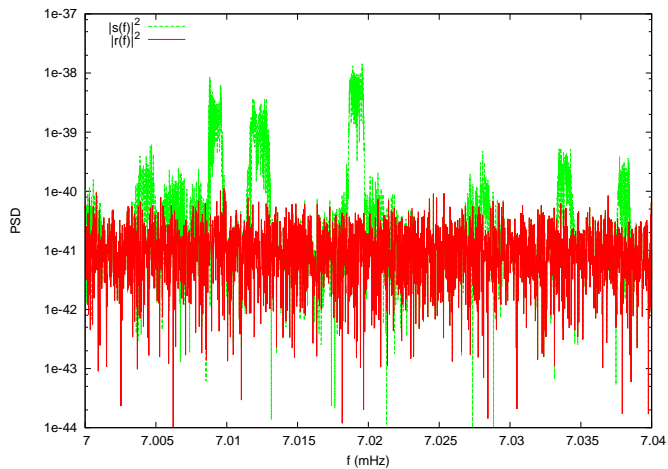} % requires the graphicx package
   \caption{Same as Figure~\ref{Algorithm:data}, but with the red [solid] line replaced by the residual power spectral density 
   after the best fit model has been regressed from the data.}
   \label{Algorithm:residual}
\end{figure}

After each iteration, if the new source is within some pre-defined number of frequency bins (depending on the characteristic bandwidth of sources in that window), than on the following iteration this template, along with any others that satisfy the closeness condition, and the new source included in the model, are allowed to vary.  Otherwise, the signals located in previous iterations are held fixed.  This keeps the number of ``active'' sources per window, per iteration, much lower than the total number of detectable binaries in that segment of data.  We are essentially searching on the residual of the data from previous iterations, but with the caveat that if new sources get too close to existing detections than all nearby signals need to be simultaneously  re-characterized.  Figure~\ref{Algorithm:residual} shows the same data segment as in Figure~\ref{Algorithm:data}, but with the best fit model regressed from the original data.  The residual is consistent with stationary 
Gaussian noise at the level of the instrument noise.

%\section{Example Results} 
%\label{example}
%\input{example.tex}

\section{The recovered source catalogue} 
\label{catalogue}
We performed a comprehensive test run of the algorithm on the MLDC challenge 4 training data set, 
containing only instrument noise and the signals from the simulated galaxy.  As this run was to serve as a test of the 
algorithm, the search was not carried to completion in all frequency windows.  In this section we 
quantify the performance of the algorithm by comparing the recovered catalogue to the source list 
supplied with the training data.  

\subsubsection{Evaluating the recovered catalogue}
The total number of detected galactic binaries that we located in the data before halting the search
was $\sim9000$.  The merit of our recovered catalogue was judged using the MLDC challenge 3 
evaluation software available as part of the {\tt lisatools} software package~\cite{lisatools}.  

\begin{figure}[htbp]
   \includegraphics[angle=270, width=0.5\textwidth]{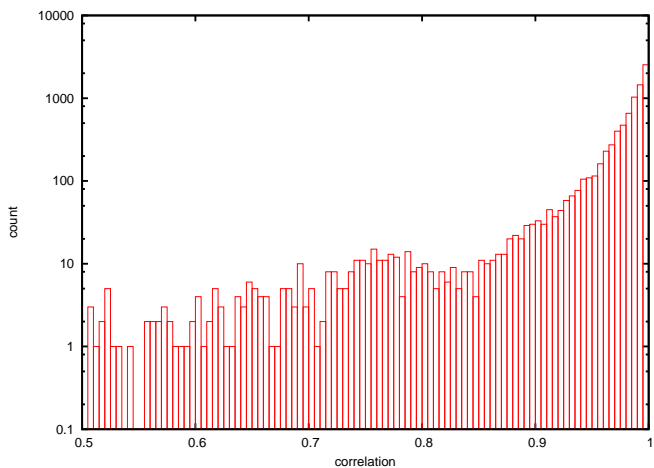} % requires the graphicx package
   \caption{Correlations between recovered sources and their corresponding signal in the training data.  Of the $\sim9000$ detections over 90$\%$ had a correlation above 0.9 with an injected signal.}
   \label{Catalogue:correlation}
\end{figure}

For each recovered source $h_{\rm rec}$, the evaluation software searches through the list of (${\rm SNR}\gtrsim1$) sources in the simulated 
galaxy, and determines which injected signal $h_{\rm inj}$ gives the lowest noise-weighted residual $(h_{\rm inj}-h_{\rm rec}|h_{\rm inj}-h_{\rm rec})$.  Once $h_{\rm rec}$ is paired with the corresponding $h_{\rm inj}$ the correlation between the two waveforms 
\begin{equation}
{\rm Corr}_{\rm inj,rec}\equiv\frac{(h_{\rm inj}|h_{\rm rec})}{\sqrt{(h_{\rm inj}|h_{\rm inj})(h_{\rm rec}|h_{\rm rec})}}
\end{equation}
is computed and stored in a ``report card'' for the recovered catalogue.  A histogram of the correlations for our recovered population of sources is depicted in Figure~\ref{Catalogue:correlation}.

\begin{figure*}[htbp]
   \includegraphics[angle=0, width=1.0\textwidth]{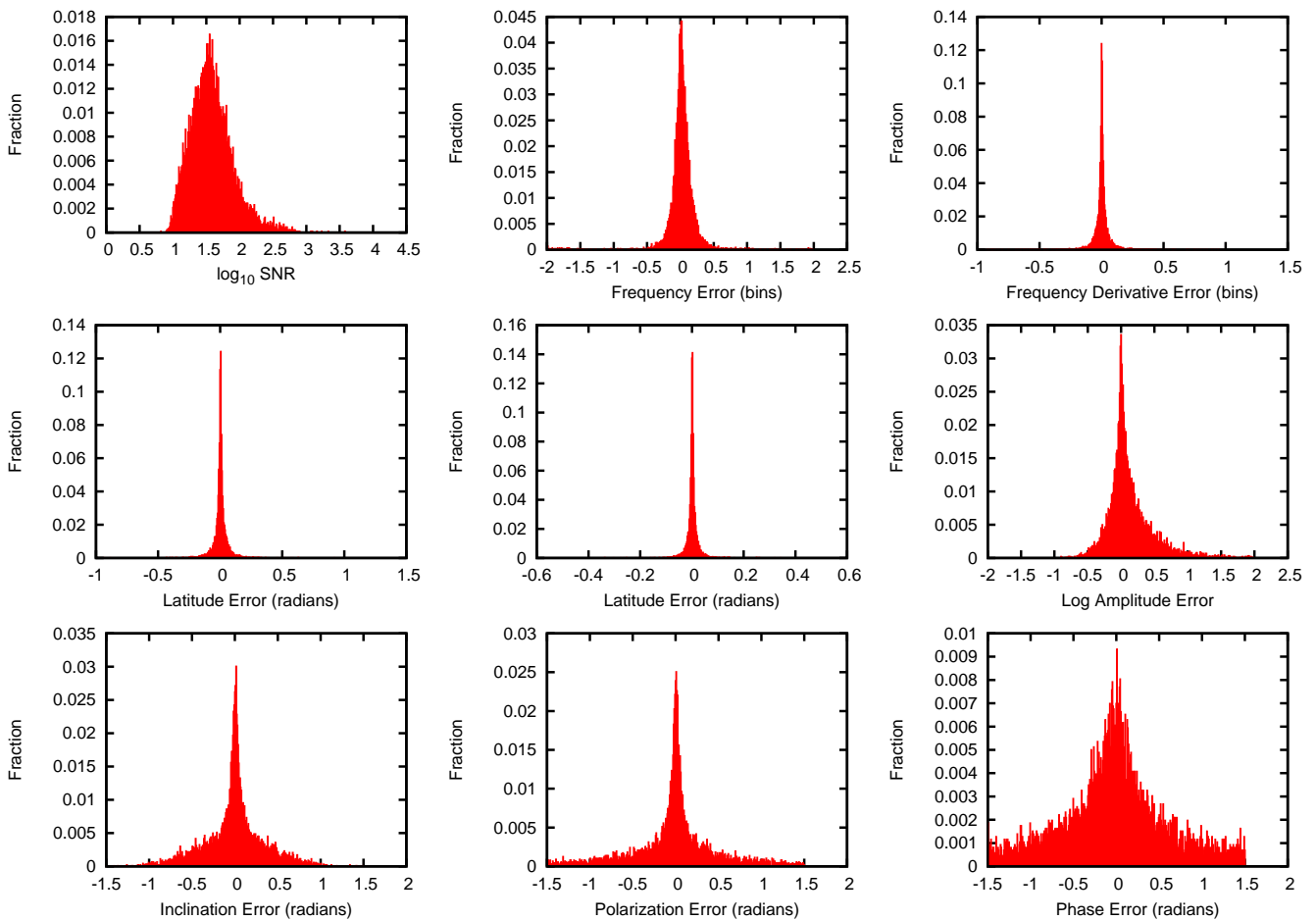} % requires the graphicx package
   \caption{[Top left]  The distribution of SNRs for the recovered signals.  [All others]  Distributions of the difference between the recovered signal parameters and the associated source 
   in the data.  We see no significant systematic biases between our best fit parameters and true values for the 
   recovered sources.  The orientation parameters of the binary are typically not well measured, hence the 
   larger tails on the error distributions.   }
   \label{Catalogue:error}
\end{figure*}

Along with the correlation files, the MLDC evaluation software saves the parameter error between each recovered 
signals and its corresponding source in the data.  The distribution of errors for the entire catalogue will reveal any 
systematic biases in the parameter recovery.  Figure~\ref{Catalogue:error} shows the distribution of recovered 
parameter biases.  We are pleased to report that all parameters show a strong peak at zero bias.  

\begin{figure}[htbp]
   \includegraphics[angle=270, width=0.5\textwidth]{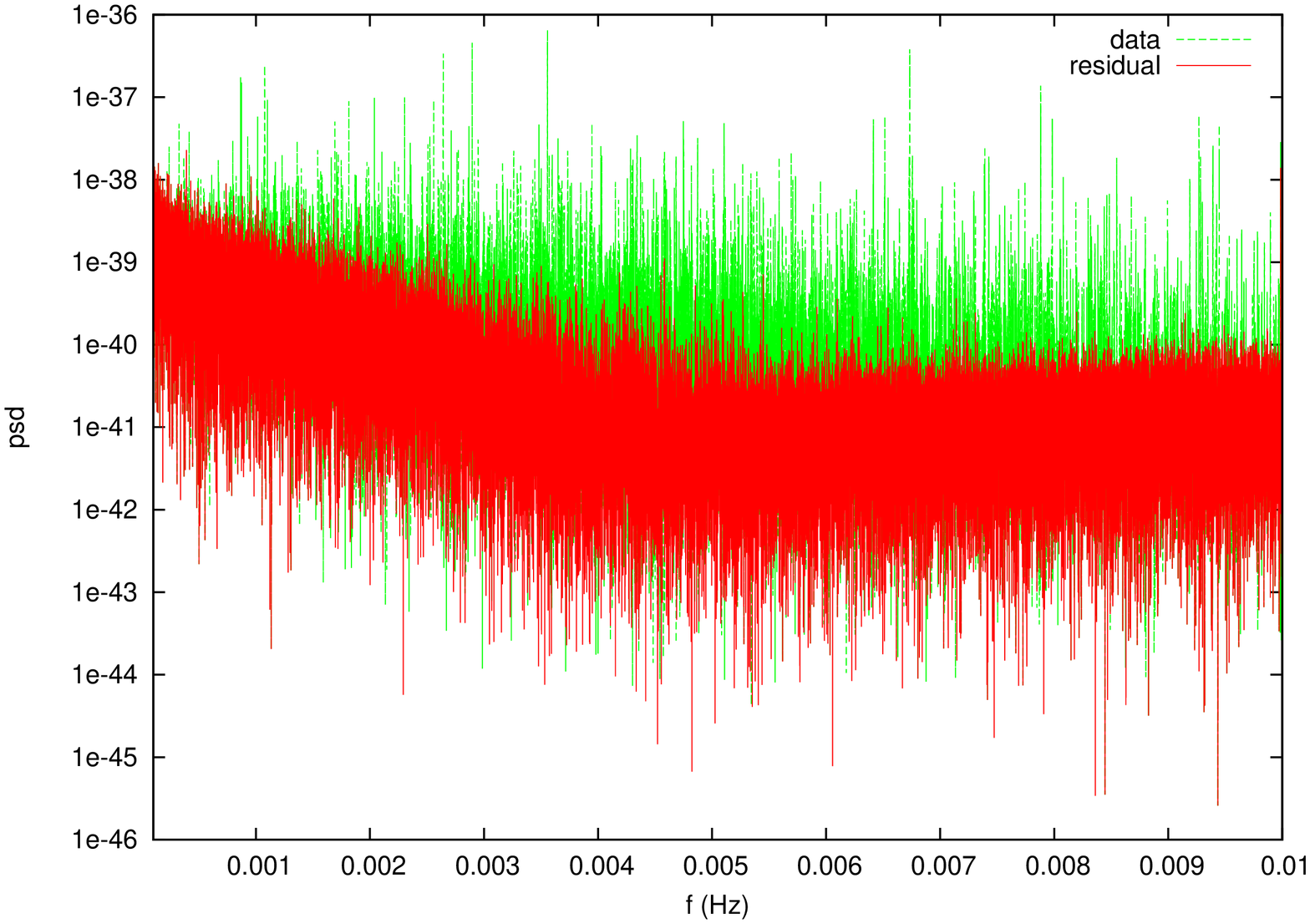} % requires the graphicx package
   \caption{Power spectral density of the training data (dashed/green) and residual (solid/red) after removal 
   of the recovered sources.  The remaining spikes in the residual are detectable signals 
   left behind after we prematurely halted the search in favor of analyzing the blind data.}
   \label{Catalogue:residual_full}
\end{figure}

The galaxy search serves the dual purpose of recovering a trove of information about the galaxy, and removing a substantial amount of foreground signal-power, facilitating the search for signals at cosmological distances.  The residual 
power spectral density of the training data after having removed the recovered signals is shown in Figure~\ref{Catalogue:residual_full}.  The dashed [green] trace shows the galaxy-only training data (without BHBs, EMRIs, etc.), while the solid [red] line is the residual after our recovered catalogue has been regressed from the data.  Notice we only performed the search out to 0.01 Hz.  At higher frequencies the signals are typically bright and isolated.  This makes them easy to find, but computationally expensive to characterize, as the bandwidth of the signal increases with frequency and amplitude.  Individual high frequency windows have been tested and will be included in our blind search, but we omit 
them from the results here.  It is clear from the residual that there are more detectable sources than the 9000 on which 
we are reporting.  However, as a test of the algorithm we take this as a satisfying result.

%\begin{figure}[htbp]
%   \includegraphics[angle=270, width=0.5\textwidth]{figures/residual_zoom.eps} % %requires the graphicx package
%   \caption{Zoom Residual}
%   \label{Catalogue:residual_zoom}
%\end{figure}

\subsubsection{Galactic binary astronomy from the recovered catalogue}
\begin{figure*}[htbp]
   \includegraphics[angle=270, width=1\textwidth]{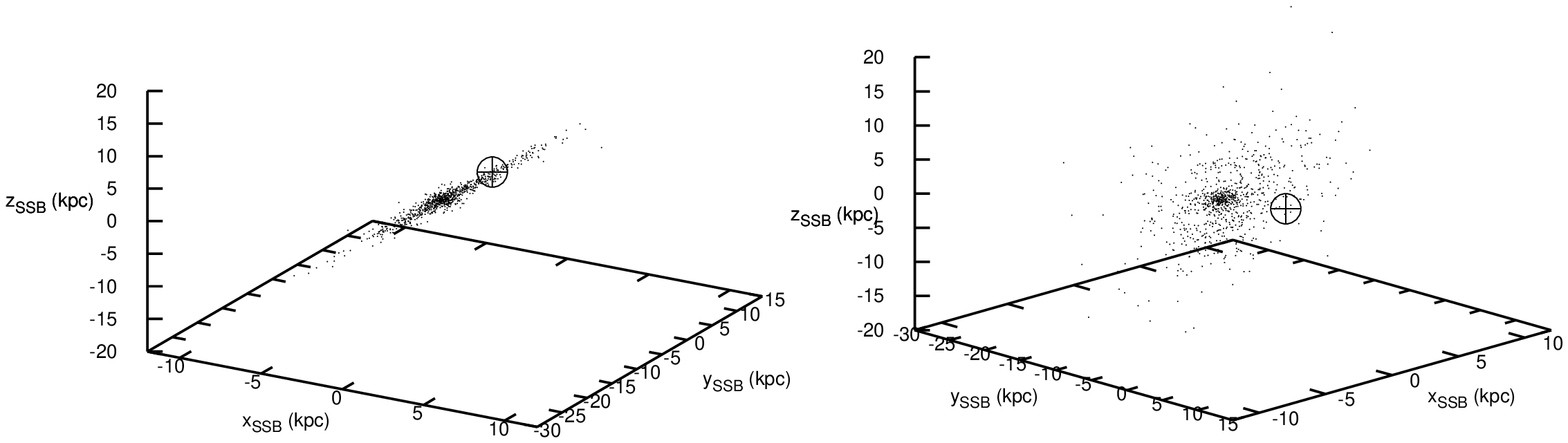} % requires the graphicx package
   \caption{Scatter plots showing the spatial distribution of recovered signals for which $D_L$ could be 
   measured in ecliptic coordinates from different viewing angles. The $\oplus$ symbol is located at the solar system barycenter.}
   \label{Catalogue:location}
\end{figure*}
%\begin{figure*}[htbp]
%   \includegraphics[angle=270, width=1\textwidth]{figures/fisher.eps} % requires the graphicx package
%   \caption{Location}
%   \label{Catalogue:fisher}
%\end{figure*}
While the focus of this work has been the algorithm and its performance on simulated data, the real fun begins 
when we brandish the source catalogue as an unprecedented astronomical tool.  Our follow-on work to this paper will 
address some specifics, but in the mean time there is one piece of ``low hanging fruit'' that is too good to pass up.  As mentioned in \S\ref{waveforms}, the inclusion of $\dot{f}_0$ in the waveform model presents us with an opportunity to 
disentangle the luminosity distance from the overall GW amplitude.  To demonstrate this capability, we select from 
the catalogue any binary with SNR$>$15 and $\dot{f}_0T^2 \geq 1$ and compute $D_L$ using Eqs.~\ref{Amp} and~\ref{fdot}.  
The resulting galactic map is shown in Figure~\ref{Catalogue:location}, exhibiting the unique capability of low frequency gravitational wave 
detectors to reconstruct a three-dimensional map of the entire Galaxy.  

Admittedly, we artificially benefit from knowing that 
any binary with $\dot{f}_0 > 0$ is dynamically evolving under radiation reaction forces only.  How this assumption biases the spatial distribution of the detected population is a top priority of our subsequent studies.

\section{Discussion}
\label{discussion}
Using~\cite{Crowder:2006eu} as a foundation, we have developed our own galactic binary detection code and tested it on 
the round 4.0 training data supplied by the Mock LISA Data Challenge Task Force.  New features to this analysis algorithm are 
the inclusion of parallel tempering and delayed rejection to increase the efficiency of the search, and using a trans-dimensional MCMC to simultaneously characterize the detected binaries and determine the most likely number of binaries 
in each small-bandwidth segment of data being analyzed.  We also include the $\dot{f}_0$ parameter in the waveform modeling which was not part of the original work by Crowder and Cornish, but was included in the MLDC round 3 data and 
subsequent entries.

The codes were written for the purpose 
of participating in the blind challenge, so the analysis of the training data was halted before completion.  
Before moving on to the challenge data, 
we recovered $\sim 
9000$ sources, 90\% of which matched one of the injected sources with correlation greater than 0.90.  It is worth noting that a 
smaller-scale space borne gravitational wave detector is more likely to fly in the near future than the full 5 Gm LISA mission 
concept.  If anything, that makes our current search software overqualified to handle whatever data we are presented.

Of course, detecting the sources is the first step towards using the data to its full potential, as a unique probe of the Galaxy.  
Including the first time derivative of the frequency in the waveform model allows us to compute the distance to binaries whose 
dynamics are not affected by mass transfer or tidal effects.  From the recovered source catalogue, we measured the sky-location and $\dot{f}$ well enough to constrain $D_L$ of $\sim 1000$ binaries, sampled from the entire volume of the Galaxy.  This result hints at the potential for low-frequency gravitational wave astronomy to offer an unprecedented view of how compact stellar remnants are distributed.  Implicit in the calculation of the distance is the supposition that all mass-transferring systems have $\dot{f}<0$.  How our ability to map the distribution of the Galaxy is impacted by relaxing this assumption will be the focus of follow-on work.  

\section{Acknowledgments}
We thank Neil Cornish and John Baker for numerous helpful suggestions and discussions, as well as Matt Adams and Bernard Kelly.  The analysis of the MLDC data was carried out using resources from the NASA Center for Computational Sciences at Goddard Space Flight Center.  This work is supported by NASA Grant 08-ATFP08-0126.

\bibliographystyle{apsrev}
\bibliography{/Volumes/Apps_and_Docs/tlittenb/papers/papers}

\begin{thebibliography}{43}
\expandafter\ifx\csname natexlab\endcsname\relax\def\natexlab#1{#1}\fi
\expandafter\ifx\csname bibnamefont\endcsname\relax
  \def\bibnamefont#1{#1}\fi
\expandafter\ifx\csname bibfnamefont\endcsname\relax
  \def\bibfnamefont#1{#1}\fi
\expandafter\ifx\csname citenamefont\endcsname\relax
  \def\citenamefont#1{#1}\fi
\expandafter\ifx\csname url\endcsname\relax
  \def\url#1{\texttt{#1}}\fi
\expandafter\ifx\csname urlprefix\endcsname\relax\def\urlprefix{URL }\fi
\providecommand{\bibinfo}[2]{#2}
\providecommand{\eprint}[2][]{\url{#2}}

\bibitem[{LSC()}]{LSC}
\emph{\bibinfo{title}{\emph{LIGO Scientific Collaboration}}},
  \bibinfo{howpublished}{\url{http://www.ligo.org/}}.

\bibitem[{Vir()}]{Virgo}
\emph{\bibinfo{title}{\emph{Virgo European Gravitational Observatory}}},
  \bibinfo{howpublished}{\url{https://www.ego-gw.it}}.

\bibitem[{\citenamefont{Stroeer and Vecchio}(2006)}]{Stroeer:2006rx}
\bibinfo{author}{\bibfnamefont{A.}~\bibnamefont{Stroeer}} \bibnamefont{and}
  \bibinfo{author}{\bibfnamefont{A.}~\bibnamefont{Vecchio}},
  \bibinfo{journal}{Class. Quant. Grav.} \textbf{\bibinfo{volume}{23}},
  \bibinfo{pages}{S809} (\bibinfo{year}{2006}), \eprint{astro-ph/0605227}.

\bibitem[{\citenamefont{Hils et~al.}(1990)\citenamefont{Hils, Bender, and
  Webbink}}]{Hils:1990vc}
\bibinfo{author}{\bibfnamefont{D.}~\bibnamefont{Hils}},
  \bibinfo{author}{\bibfnamefont{P.~L.} \bibnamefont{Bender}},
  \bibnamefont{and} \bibinfo{author}{\bibfnamefont{R.~F.}
  \bibnamefont{Webbink}}, \bibinfo{journal}{Astrophys. J.}
  \textbf{\bibinfo{volume}{360}}, \bibinfo{pages}{75} (\bibinfo{year}{1990}).

\bibitem[{\citenamefont{Nelemans et~al.}(2001)\citenamefont{Nelemans,
  Yungelson, and Portegies~Zwart}}]{Nelemans:2001hp}
\bibinfo{author}{\bibfnamefont{G.}~\bibnamefont{Nelemans}},
  \bibinfo{author}{\bibfnamefont{L.~R.} \bibnamefont{Yungelson}},
  \bibnamefont{and} \bibinfo{author}{\bibfnamefont{S.~F.}
  \bibnamefont{Portegies~Zwart}}, \bibinfo{journal}{Astron. Astrophys.}
  \textbf{\bibinfo{volume}{375}}, \bibinfo{pages}{890} (\bibinfo{year}{2001}),
  \eprint{astro-ph/0105221}.

\bibitem[{\citenamefont{Nelemans et~al.}(2004)\citenamefont{Nelemans,
  Yungelson, and Portegies~Zwart}}]{Nelemans:2003ha}
\bibinfo{author}{\bibfnamefont{G.}~\bibnamefont{Nelemans}},
  \bibinfo{author}{\bibfnamefont{L.~R.} \bibnamefont{Yungelson}},
  \bibnamefont{and} \bibinfo{author}{\bibfnamefont{S.~F.}
  \bibnamefont{Portegies~Zwart}}, \bibinfo{journal}{Mon. Not. Roy. Astron.
  Soc.} \textbf{\bibinfo{volume}{349}}, \bibinfo{pages}{181}
  (\bibinfo{year}{2004}), \eprint{astro-ph/0312193}.

\bibitem[{\citenamefont{Bender et~al.}(1998)}]{LISA}
\bibinfo{author}{\bibfnamefont{P.}~\bibnamefont{Bender}} \bibnamefont{et~al.},
  \bibinfo{journal}{LISA Pre-Phase A Report}  (\bibinfo{year}{1998}).

\bibitem[{MLD()}]{MLDC}
\emph{\bibinfo{title}{\emph{Mock LISA Data Challenge}}},
  \bibinfo{howpublished}{\url{http://astrogravs.nasa.gov/docs/mldc}}.

\bibitem[{\citenamefont{Babak et~al.}(2008)}]{Babak:2008sn}
\bibinfo{author}{\bibfnamefont{S.}~\bibnamefont{Babak}} \bibnamefont{et~al.},
  \bibinfo{journal}{Class. Quant. Grav.} \textbf{\bibinfo{volume}{25}},
  \bibinfo{pages}{184026} (\bibinfo{year}{2008}), \eprint{0806.2110}.

\bibitem[{\citenamefont{Crowder and Cornish}(2007)}]{Crowder:2006eu}
\bibinfo{author}{\bibfnamefont{J.}~\bibnamefont{Crowder}} \bibnamefont{and}
  \bibinfo{author}{\bibfnamefont{N.}~\bibnamefont{Cornish}},
  \bibinfo{journal}{Phys. Rev.} \textbf{\bibinfo{volume}{D75}},
  \bibinfo{pages}{043008} (\bibinfo{year}{2007}), \eprint{astro-ph/0611546}.

\bibitem[{\citenamefont{Blaut et~al.}(2010)\citenamefont{Blaut, Babak, and
  Krolak}}]{Blaut:2009si}
\bibinfo{author}{\bibfnamefont{A.}~\bibnamefont{Blaut}},
  \bibinfo{author}{\bibfnamefont{S.}~\bibnamefont{Babak}}, \bibnamefont{and}
  \bibinfo{author}{\bibfnamefont{A.}~\bibnamefont{Krolak}},
  \bibinfo{journal}{Phys. Rev.} \textbf{\bibinfo{volume}{D81}},
  \bibinfo{pages}{063008} (\bibinfo{year}{2010}), \eprint{0911.3020}.

\bibitem[{\citenamefont{Whelan et~al.}(2010)\citenamefont{Whelan, Prix, and
  Khurana}}]{Whelan:2009jr}
\bibinfo{author}{\bibfnamefont{J.~T.} \bibnamefont{Whelan}},
  \bibinfo{author}{\bibfnamefont{R.}~\bibnamefont{Prix}}, \bibnamefont{and}
  \bibinfo{author}{\bibfnamefont{D.}~\bibnamefont{Khurana}},
  \bibinfo{journal}{Class. Quant. Grav.} \textbf{\bibinfo{volume}{27}},
  \bibinfo{pages}{055010} (\bibinfo{year}{2010}), \eprint{0908.3766}.

\bibitem[{\citenamefont{Trias et~al.}(2009)\citenamefont{Trias, Vecchio, and
  Veitch}}]{Trias:2009km}
\bibinfo{author}{\bibfnamefont{M.}~\bibnamefont{Trias}},
  \bibinfo{author}{\bibfnamefont{A.}~\bibnamefont{Vecchio}}, \bibnamefont{and}
  \bibinfo{author}{\bibfnamefont{J.}~\bibnamefont{Veitch}},
  \bibinfo{journal}{Class. Quant. Grav.} \textbf{\bibinfo{volume}{26}},
  \bibinfo{pages}{204024} (\bibinfo{year}{2009}), \eprint{0905.2976}.

\bibitem[{\citenamefont{Umstatter
  et~al.}(2005{\natexlab{a}})}]{Umstatter:2005jd}
\bibinfo{author}{\bibfnamefont{R.}~\bibnamefont{Umstatter}}
  \bibnamefont{et~al.}, \bibinfo{journal}{Phys. Rev.}
  \textbf{\bibinfo{volume}{D72}}, \bibinfo{pages}{022001}
  (\bibinfo{year}{2005}{\natexlab{a}}), \eprint{gr-qc/0506055}.

\bibitem[{\citenamefont{Littenberg and Cornish}(2009)}]{Littenberg:2009bm}
\bibinfo{author}{\bibfnamefont{T.~B.} \bibnamefont{Littenberg}}
  \bibnamefont{and} \bibinfo{author}{\bibfnamefont{N.~J.}
  \bibnamefont{Cornish}}, \bibinfo{journal}{Phys. Rev.}
  \textbf{\bibinfo{volume}{D80}}, \bibinfo{pages}{063007}
  (\bibinfo{year}{2009}), \eprint{0902.0368}.

\bibitem[{\citenamefont{Stroeer and Veitch}(2009)}]{Stroeer:2009hj}
\bibinfo{author}{\bibfnamefont{A.}~\bibnamefont{Stroeer}} \bibnamefont{and}
  \bibinfo{author}{\bibfnamefont{J.}~\bibnamefont{Veitch}},
  \bibinfo{journal}{Phys. Rev.} \textbf{\bibinfo{volume}{D80}},
  \bibinfo{pages}{064032} (\bibinfo{year}{2009}), \eprint{0907.2198}.

\bibitem[{\citenamefont{Gregory}(2006)}]{Gregory}
\bibinfo{author}{\bibfnamefont{P.~C.} \bibnamefont{Gregory}},
  \emph{\bibinfo{title}{Bayesian Logical Data Analysis for the Physical
  Sciences}} (\bibinfo{publisher}{Cambridge University Press},
  \bibinfo{year}{2006}).

\bibitem[{\citenamefont{Veitch and Vecchio}(2008)}]{Veitch:2008ur}
\bibinfo{author}{\bibfnamefont{J.}~\bibnamefont{Veitch}} \bibnamefont{and}
  \bibinfo{author}{\bibfnamefont{A.}~\bibnamefont{Vecchio}},
  \bibinfo{journal}{Phys. Rev.} \textbf{\bibinfo{volume}{D78}},
  \bibinfo{pages}{022001} (\bibinfo{year}{2008}), \eprint{0801.4313}.

\bibitem[{\citenamefont{Feroz et~al.}(2009)\citenamefont{Feroz, Gair, Hobson,
  and Porter}}]{Feroz:2009de}
\bibinfo{author}{\bibfnamefont{F.}~\bibnamefont{Feroz}},
  \bibinfo{author}{\bibfnamefont{J.~R.} \bibnamefont{Gair}},
  \bibinfo{author}{\bibfnamefont{M.~P.} \bibnamefont{Hobson}},
  \bibnamefont{and} \bibinfo{author}{\bibfnamefont{E.~K.} \bibnamefont{Porter}}
  (\bibinfo{year}{2009}), \eprint{0904.1544}.

\bibitem[{\citenamefont{Swendsen and Wang}(1986)}]{Swendsen:1986}
\bibinfo{author}{\bibfnamefont{R.}~\bibnamefont{Swendsen}} \bibnamefont{and}
  \bibinfo{author}{\bibfnamefont{J.}~\bibnamefont{Wang}},
  \bibinfo{journal}{Phys. Rev. Lett.} \textbf{\bibinfo{volume}{57}},
  \bibinfo{pages}{2607} (\bibinfo{year}{1986}).

\bibitem[{\citenamefont{van~der Sluys et~al.}(2008)}]{vanderSluys:2008qx}
\bibinfo{author}{\bibfnamefont{M.}~\bibnamefont{van~der Sluys}}
  \bibnamefont{et~al.}, \bibinfo{journal}{Class. Quant. Grav.}
  \textbf{\bibinfo{volume}{25}}, \bibinfo{pages}{184011}
  (\bibinfo{year}{2008}), \eprint{0805.1689}.

\bibitem[{\citenamefont{Littenberg and Cornish}(2010)}]{Littenberg:2010gf}
\bibinfo{author}{\bibfnamefont{T.~B.} \bibnamefont{Littenberg}}
  \bibnamefont{and} \bibinfo{author}{\bibfnamefont{N.~J.}
  \bibnamefont{Cornish}}, \bibinfo{journal}{Phys. Rev.}
  \textbf{\bibinfo{volume}{D82}}, \bibinfo{pages}{103007}
  (\bibinfo{year}{2010}), \eprint{1008.1577}.

\bibitem[{\citenamefont{Cornish et~al.}(2011)\citenamefont{Cornish, Sampson,
  Yunes, and Pretorius}}]{Cornish:2011ys}
\bibinfo{author}{\bibfnamefont{N.}~\bibnamefont{Cornish}},
  \bibinfo{author}{\bibfnamefont{L.}~\bibnamefont{Sampson}},
  \bibinfo{author}{\bibfnamefont{N.}~\bibnamefont{Yunes}}, \bibnamefont{and}
  \bibinfo{author}{\bibfnamefont{F.}~\bibnamefont{Pretorius}}
  (\bibinfo{year}{2011}), \eprint{1105.2088}.

\bibitem[{\citenamefont{Green}(1995)}]{Green:1995}
\bibinfo{author}{\bibfnamefont{P.}~\bibnamefont{Green}},
  \bibinfo{journal}{Biometrika} \textbf{\bibinfo{volume}{82}},
  \bibinfo{pages}{711} (\bibinfo{year}{1995}).

\bibitem[{\citenamefont{Green}(2003)}]{Green:2003}
\bibinfo{author}{\bibfnamefont{P.}~\bibnamefont{Green}},
  \emph{\bibinfo{title}{Highly Structured Stochastic Systems}}
  (\bibinfo{publisher}{Oxford University Press}, \bibinfo{year}{2003}).

\bibitem[{\citenamefont{Umstatter
  et~al.}(2005{\natexlab{b}})}]{Umstatter:2005su}
\bibinfo{author}{\bibfnamefont{R.}~\bibnamefont{Umstatter}}
  \bibnamefont{et~al.}, \bibinfo{journal}{Class. Quant. Grav.}
  \textbf{\bibinfo{volume}{22}}, \bibinfo{pages}{S901}
  (\bibinfo{year}{2005}{\natexlab{b}}), \eprint{gr-qc/0503121}.

\bibitem[{\citenamefont{Tinto et~al.}(2002)\citenamefont{Tinto, Estabrook, and
  Armstrong}}]{Tinto:2002de}
\bibinfo{author}{\bibfnamefont{M.}~\bibnamefont{Tinto}},
  \bibinfo{author}{\bibfnamefont{F.~B.} \bibnamefont{Estabrook}},
  \bibnamefont{and} \bibinfo{author}{\bibfnamefont{J.~W.}
  \bibnamefont{Armstrong}}, \bibinfo{journal}{Phys. Rev.}
  \textbf{\bibinfo{volume}{D65}}, \bibinfo{pages}{082003}
  (\bibinfo{year}{2002}).

\bibitem[{\citenamefont{Stroeer et~al.}(2005)\citenamefont{Stroeer, Vecchio,
  and Nelemans}}]{Stroeer:2005cv}
\bibinfo{author}{\bibfnamefont{A.}~\bibnamefont{Stroeer}},
  \bibinfo{author}{\bibfnamefont{A.}~\bibnamefont{Vecchio}}, \bibnamefont{and}
  \bibinfo{author}{\bibfnamefont{G.}~\bibnamefont{Nelemans}},
  \bibinfo{journal}{Astrophys. J.} \textbf{\bibinfo{volume}{633}},
  \bibinfo{pages}{L33} (\bibinfo{year}{2005}), \eprint{astro-ph/0509632}.

\bibitem[{\citenamefont{Willems et~al.}(2008)\citenamefont{Willems, Vecchio,
  and Kalogera}}]{Willems:2007nq}
\bibinfo{author}{\bibfnamefont{B.}~\bibnamefont{Willems}},
  \bibinfo{author}{\bibfnamefont{A.}~\bibnamefont{Vecchio}}, \bibnamefont{and}
  \bibinfo{author}{\bibfnamefont{V.}~\bibnamefont{Kalogera}},
  \bibinfo{journal}{Phys. Rev. Lett.} \textbf{\bibinfo{volume}{100}},
  \bibinfo{pages}{041102} (\bibinfo{year}{2008}), \eprint{0706.3700}.

\bibitem[{\citenamefont{Stroeer and Nelemans}(2009)}]{Stroeer:2009uy}
\bibinfo{author}{\bibfnamefont{A.}~\bibnamefont{Stroeer}} \bibnamefont{and}
  \bibinfo{author}{\bibfnamefont{G.}~\bibnamefont{Nelemans}}
  (\bibinfo{year}{2009}), \eprint{0909.1796}.

\bibitem[{\citenamefont{Larson and Littenberg}(2011)}]{Larson}
\bibinfo{author}{\bibfnamefont{S.~L.} \bibnamefont{Larson}} \bibnamefont{and}
  \bibinfo{author}{\bibfnamefont{T.~B.} \bibnamefont{Littenberg}},
  \bibinfo{journal}{in preperation}  (\bibinfo{year}{2011}).

\bibitem[{\citenamefont{Cornish and Littenberg}(2007)}]{Cornish:2007if}
\bibinfo{author}{\bibfnamefont{N.~J.} \bibnamefont{Cornish}} \bibnamefont{and}
  \bibinfo{author}{\bibfnamefont{T.~B.} \bibnamefont{Littenberg}},
  \bibinfo{journal}{Phys. Rev.} \textbf{\bibinfo{volume}{D76}},
  \bibinfo{pages}{083006} (\bibinfo{year}{2007}), \eprint{0704.1808}.

\bibitem[{\citenamefont{Adams}(2011)}]{Adams_private}
\bibinfo{author}{\bibfnamefont{M.~R.} \bibnamefont{Adams}},
  \bibinfo{journal}{private communication}  (\bibinfo{year}{2011}).

\bibitem[{\citenamefont{Babak et~al.}(2010)}]{Babak:2009cj}
\bibinfo{author}{\bibfnamefont{S.}~\bibnamefont{Babak}} \bibnamefont{et~al.}
  (\bibinfo{collaboration}{Mock LISA Data Challenge Task Force}),
  \bibinfo{journal}{Class. Quant. Grav.} \textbf{\bibinfo{volume}{27}},
  \bibinfo{pages}{084009} (\bibinfo{year}{2010}), \eprint{0912.0548}.

\bibitem[{\citenamefont{Jaranowski et~al.}(1998)\citenamefont{Jaranowski,
  Krolak, and Schutz}}]{Jaranowski:1998qm}
\bibinfo{author}{\bibfnamefont{P.}~\bibnamefont{Jaranowski}},
  \bibinfo{author}{\bibfnamefont{A.}~\bibnamefont{Krolak}}, \bibnamefont{and}
  \bibinfo{author}{\bibfnamefont{B.~F.} \bibnamefont{Schutz}},
  \bibinfo{journal}{Phys. Rev.} \textbf{\bibinfo{volume}{D58}},
  \bibinfo{pages}{063001} (\bibinfo{year}{1998}), \eprint{gr-qc/9804014}.

\bibitem[{\citenamefont{Krolak et~al.}(2004)\citenamefont{Krolak, Tinto, and
  Vallisneri}}]{Krolak:2004xp}
\bibinfo{author}{\bibfnamefont{A.}~\bibnamefont{Krolak}},
  \bibinfo{author}{\bibfnamefont{M.}~\bibnamefont{Tinto}}, \bibnamefont{and}
  \bibinfo{author}{\bibfnamefont{M.}~\bibnamefont{Vallisneri}},
  \bibinfo{journal}{Phys. Rev.} \textbf{\bibinfo{volume}{D70}},
  \bibinfo{pages}{022003} (\bibinfo{year}{2004}), \eprint{gr-qc/0401108}.

\bibitem[{\citenamefont{Cutler and Schutz}(2005)}]{Cutler:2005hc}
\bibinfo{author}{\bibfnamefont{C.}~\bibnamefont{Cutler}} \bibnamefont{and}
  \bibinfo{author}{\bibfnamefont{B.~F.} \bibnamefont{Schutz}},
  \bibinfo{journal}{Phys. Rev.} \textbf{\bibinfo{volume}{D72}},
  \bibinfo{pages}{063006} (\bibinfo{year}{2005}), \eprint{gr-qc/0504011}.

\bibitem[{\citenamefont{Cornish and Porter}(2007)}]{Cornish:2006ms}
\bibinfo{author}{\bibfnamefont{N.~J.} \bibnamefont{Cornish}} \bibnamefont{and}
  \bibinfo{author}{\bibfnamefont{E.~K.} \bibnamefont{Porter}},
  \bibinfo{journal}{Class. Quant. Grav.} \textbf{\bibinfo{volume}{24}},
  \bibinfo{pages}{5729} (\bibinfo{year}{2007}), \eprint{gr-qc/0612091}.

\bibitem[{\citenamefont{Tierney and Mira}(1999)}]{Tierney:1999}
\bibinfo{author}{\bibfnamefont{L.}~\bibnamefont{Tierney}} \bibnamefont{and}
  \bibinfo{author}{\bibfnamefont{A.}~\bibnamefont{Mira}},
  \bibinfo{journal}{Stat. Med.} \textbf{\bibinfo{volume}{18}}
  (\bibinfo{year}{1999}).

\bibitem[{\citenamefont{Mira}(2001)}]{Mira:2001}
\bibinfo{author}{\bibfnamefont{A.}~\bibnamefont{Mira}},
  \bibinfo{journal}{Metron} \textbf{\bibinfo{volume}{LIX}},
  \bibinfo{pages}{231} (\bibinfo{year}{2001}).

\bibitem[{\citenamefont{Farr and Mandel}(2011)}]{Farr:2011sk}
\bibinfo{author}{\bibfnamefont{W.~M.} \bibnamefont{Farr}} \bibnamefont{and}
  \bibinfo{author}{\bibfnamefont{I.}~\bibnamefont{Mandel}}
  (\bibinfo{year}{2011}), \eprint{1104.0984}.

\bibitem[{\citenamefont{Raftery}(1996)}]{Raftery:1996}
\bibinfo{author}{\bibfnamefont{A.}~\bibnamefont{Raftery}},
  \emph{\bibinfo{title}{Practical Markov Chain Monte Carlo}}
  (\bibinfo{publisher}{Chapman and Hall}, \bibinfo{address}{London},
  \bibinfo{year}{1996}).

\bibitem[{lis()}]{lisatools}
\emph{\bibinfo{title}{\emph{lisatools}}},
  \bibinfo{howpublished}{\url{http://code.google.com/p/lisatools}}.

\end{thebibliography}

\end{document}